# A coarse-grained Langevin molecular dynamics approach to *de novo* protein structure prediction

Takeshi N. Sasaki, Hikmet Cetin and Masaki Sasai

Department of Computational Science and Engineering, Nagoya University, Nagoya 464-8603, Japan

ABSTRACT

*De novo* prediction of protein structures, the prediction of structures from amino-acid sequences which are not similar to those of hitherto resolved structures, has been one of the major challenges in molecular biophysics. In this paper, we develop a new method of *de novo* prediction, which combines the fragment assembly method and the simulation of physical folding process: Structures which have consistently assembled fragments are dynamically searched by Langevin molecular dynamics of conformational change. The benchmarking test shows that the prediction is improved when the candidate structures are cross-checked by an empirically derived score function.

KEY WORDS

protein structure prediction, Langevin dynamics, fragment assembly

# INTRODUCTION

Prediction of protein structure from amino-acid sequence is a major challenge in biophysics. As the number of determined structures increases, fairly precise prediction has become possible if the sequence of the target protein is close to the sequence of a known structure [1]. Such prediction utilizing homologous proteins is called template based modeling (TBM). For targets whose sequences do not resemble those of hitherto resolved structures, however, the prediction becomes a harder problem [2], which is known as *de novo* prediction or template free modeling (FM). It is important to develop a reliable *de novo* prediction technique not only to solve previously unseen structures but also to understand the principles of structure formation. In recent experiment of the 7th critical assessment of techniques for protein structure prediction (CASP7), results of both TBM and FM have been intensively discussed [3]. From this discussion, we can see that we still do not have a reliable consistent technique for *de novo* prediction in spite of the much effort devoted to this problem [4-12].

Following Anfinsen's thermodynamic hypothesis [13], native structures should have low free energy. In *de novo* prediction, many research groups have developed sampling techniques to find such low free energy structures by applying various types of effective energy functions. Relatively successful methods among them are the fragment assembly method [4-8] and the Threading/Assembly/Refinement (TASSER) method [9-12], which have employed the strategy to assemble the candidates of local structures such as 9-residue length fragments [4-8] or longer chain configurations [9-12]. In these methods, local structural candidates are selected at first by utilizing the local sequential similarity between target and database proteins, and then the whole chain structure is predicted by finding the consistent combination of local structural candidates to form the whole structure of the low effective energy. Success of these methods implies that consistency [14] and minimal frustration [15] among local and whole structures are the guidelines for structural formation in proteins.

Another strategy for *de novo* prediction is to use Monte Carlo [16-18] or Langevin molecular dynamics (MD) methods [19-22] to simulate the folding process. Merits of simulating physical folding process are in multiple ways. First, the method developed in the prediction problem should give insights on folding process, second, the method could be applied outside of the prediction problem to the large scale conformational change in protein functioning, and last but not least, the structure generation mimicking the process existing in nature should be a reasonable way to resolve the complex conformation.

In the present paper we discuss a newly developed *de novo* prediction method which incorporates both of above two strategies at the same time. In this method a coarse-grained energy function consisting of several terms of potentials is constructed. Some of those potentials express structural tendency for fragments to take in the target protein, and other multi-residue potentials express how the fragments are assembled through hydrophobic interactions and hydrogen-bonding. In this way, both the local structure prediction and the minimally frustrated assembly of local structures should be realized at the same time when this total energy function is lowered enough. Using thus defined energy function, Langevin MD simulations are performed to search structures of low energy. A benchmarking test of this method is performed by targeting proteins used in the TBM and FM categories of CASP7.

**METHODS**

Peptide chain is expressed by the connected beads of α carbons, whose coordinates are denoted by $\{\mathbf{r}_i\}$. Folding of a peptide chain is simulated by numerically solving an overdamped Langevin equation,

$$d\mathbf{r}_i/dt = -\partial V_{\text{total}}/\partial \mathbf{r}_i + \xi_i(t), \qquad (1)$$

where $\xi_i(t)$ is a Gaussian white noise satisfying $\langle \xi_i(t)\xi_j(t')\rangle = 2T\delta_{ij}\delta(t-t')$, and $T$ is a temperature-like parameter to control the amplitude of noise. $V_{\text{total}}$ is the multi-body potential which is an explicitly differentiable function of $\{\mathbf{r}_i\}$ having the form,

$$V_{\text{total}} = V_{\text{fragment}} + V_{\text{assemble}}, \qquad (2)$$

where $V_{\text{fragment}}$ represents interactions to form local fragmental structures and $V_{\text{assemble}}$ represents interactions to assemble fragments. $V_{\text{fragment}} = w_1 V_{\text{fragment}}^{\text{pair}} + w_2 V_{\text{fragment}}^{\text{angle}}$ and $V_{\text{assemble}} = w_3 V_{\text{nn}} + w_4 V_\beta$, where $w_1$, $w_2$, $w_3$, and $w_4$ are weight coefficients. $V_{\text{fragment}}^{\text{pair}}$, $V_{\text{fragment}}^{\text{angle}}$, and $V_{\text{nn}}$ are constructed by estimating the statistical tendency of candidate structures for fragments which are selected from the library of non-redundant protein structures.

**Fragment selection.** Each 9-residue window in sequence of the target protein is compared with fragments in the library of 3624 non-redundant protein structures, which were selected from the PDB list released before CASP7 by using PISCES server [23, 24] with its default parameters. Sequence profiles for the target protein and for each protein of the structure library are derived from the non-redundant (NR) sequence database [25] by 3 times iteration of the Position Specific Iterative (PSI)-BLAST [26] with the E-value cutoff of 0.001. For every 9-residue window in sequence of the target protein, fragments are selected from the structure library according to their profile correlation to the window sequence. These most correlated fragments selected for the $j$th 9-residue window in the target are denoted by $F_i(j)$, with $i = 1, ..., N^{\text{fragment}}$ and $N^{\text{fragment}} = 20$ or 40.

**Fragment-based two-body potential.** $V_{\text{fragment}}^{\text{pair}}$ is constructed from $\{F_i(j)\}$. When the $j$th window spans from the $j$th residue to the $j+8$th residue of the target, distance between the $j+4$th residue and the other $j + 4 \pm k$ th residue in the window with $k = 1, 2, 3$, and 4, is $r_{j+4, j+4\pm k} = |\mathbf{r}_{j+4} - \mathbf{r}_{j+4\pm k}|$. $V^{j, \pm k}(r_{j+4, j+4\pm k})$ is defined to express the statistical tendency that $r_{j+4, j+4\pm k}$ takes;

$$V^{j, \pm k}(r_{j+4, j+4\pm k}) = \frac{-1}{\sum_{i=1}^{N^{\text{fragment}}} C(i,j)} \sum_{i=1}^{N^{\text{fragment}}} C(i,j) \times \exp\left[-\frac{1}{2c_k}\left(r_{j+4, j+4\pm k} - r_{5, 5\pm k}^i\right)^2\right], \quad (3)$$

where $c_k = 0.5k^{2/3}$ and $N^{\text{fragment}} = 20$. $C(i,j)$ is the correlation coefficient between the sequence profile of the $j$th 9-residue window of the target and the sequence profile of $F_i(j)$. $r_{5, 5\pm k}^i$ is the distance between the 5th residue from the N-end of the fragment, and the $5 \pm k$ th residue in $F_i(j)$. The envelope of thus summed Gaussians represents the statistical constraint to distances in the 9-residue fragment in the target structure. See Figure 1 for an illustrative explanation of $V^{j, \pm k}(r_{j+4, j+4 \pm k})$. $V_{\text{fragment}}^{\text{pair}}$ is a sum of $V^{j, \pm k}(r_{j+4, j+4 \pm k})$ for $k$ and for all 9-residue windows,

$$V_{\text{fragment}}^{\text{pair}} = \sum_{j=1}^{N^{\text{res}}-8} \left( \sum_{k=1}^{4} V^{j,-k}(r_{j+4, j+4-k}) + \sum_{k=1}^{4} V^{j,k}(r_{j+4, j+4+k}) \right), \quad (4)$$

where $N^{\text{res}}$ is the number of residues of the target protein.

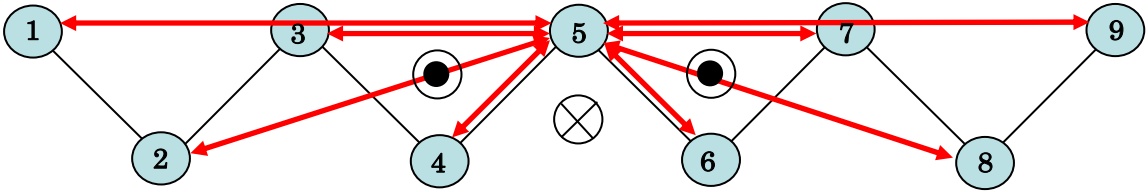

Figure 1

Illustration to explain the construction of $V_{\text{fragment}}^{\text{pair}}$ and $V_{\text{fragment}}^{\text{angle}}$. $V_{\text{fragment}}^{\text{pair}}$ describes constraints for the pair distances designated by red arrows in the fragment. $V_{\text{fragment}}^{\text{angle}}$ gives constraints for dihedral angles between three successive planes at around the center of the fragment.

**Fragment-based pseudo-dihedral angle potential.** $V_{\text{fragment}}^{\text{angle}}$ is constructed from $\{F_i(j)\}$. Two pseudo-dihedral angles, $\theta_j$ and $\phi_j$, are defined by angles between the plane containing the three points, $\mathbf{r}_j$, $\mathbf{r}_{j+1}$, and $\mathbf{r}_{j+2}$, and the plane containing $\mathbf{r}_{j+1}$, $\mathbf{r}_{j+2}$, and $\mathbf{r}_{j+3}$ as

$$\cos\theta_j = \frac{\mathbf{g}_j \cdot \mathbf{g}_{j+1}}{|\mathbf{g}_j||\mathbf{g}_{j+1}|}, \quad \cos\phi_j = \frac{\mathbf{g}_j \cdot \mathbf{r}_{j+2,j+3}}{|\mathbf{g}_j||\mathbf{r}_{j+2,j+3}|}, \tag{5}$$

where $\mathbf{g}_j = \mathbf{r}_{j,j+1} \times \mathbf{r}_{j+1,j+2}$ and $\mathbf{r}_{j,j+1} = \mathbf{r}_{j+1} - \mathbf{r}_j$. Statistical tendency of pseudo-dihedral angles at around the center of the $j$th 9-residue window of the target is expressed by,

$$V_\theta^j(\cos\theta_{j+2},\cos\theta_{j+3}) = \frac{-1}{\sum_{i=1}^{N^{\text{fragment}}} C(i,j)} \\ \times \sum_{i=1}^{N^{\text{fragment}}} C(i,j) \times \exp\left[-\frac{1}{2C_{\text{angle}}}\left\{\left(\cos\theta_{j+2} - \cos\theta_3^i\right)^2 + \left(\cos\theta_{j+3} - \cos\theta_4^i\right)^2\right\}\right], \tag{6}$$

$$V_\phi^j(\cos\phi_{j+2},\cos\phi_{j+3}) = \frac{-1}{\sum_{i=1}^{N^{\text{fragment}}} C(i,j)} \\ \times \sum_{i=1}^{N^{\text{fragment}}} C(i,j) \times \exp\left[-\frac{1}{2C_{\text{angle}}}\left\{\left(\cos\phi_{j+2} - \cos\phi_3^i\right)^2 + \left(\cos\phi_{j+3} - \cos\phi_4^i\right)^2\right\}\right], \tag{7}$$

where $C_{\text{angle}} = 0.05$, $N^{\text{fragment}} = 20$, and $\theta_3^i, \theta_4^i, \phi_3^i,$ and $\phi_4^i$ are the pseudo-dihedral angles defined by the five consecutive residues around the center of $F_i(j)$. As illustrated in Figure 1, these potentials provide geometrical constraints to five consecutive residues from $j+2$ to $j+6$ in the $j$th window of the target protein. $V_{\text{fragment}}^{\text{angle}}$ is a sum of

$V_\theta^j(\cos\theta_{j+2}, \cos\theta_{j+3})$ and $V_\phi^j(\cos\phi_{j+2}, \cos\phi_{j+3})$ for all the 9-residue windows in the target;

$$V_{\text{fragment}}^{\text{angle}} = \sum_{j=1}^{N^{res}-8} \left(V_\theta^j + V_\phi^j\right) . \tag{8}$$

**Neighboring-number potential.** $V_{nn}$ expresses the hydrophobic interaction and the steric exclusive repulsion. Around the center residue, *i.e.* the *j*+4th residue in the *j*th window in the target protein, spheres of radius $r(k) = 2.0 + 2.0 \times k$ (Å) with $k = 1, 2, 3,$ and $4$ are defined as shown in Figure 2. The number of neighboring residues around the center residue is counted by defining a smooth differentiable function,

$$u_k(r) = \begin{cases} 1 & (r < r(k) - \delta r) \\ \frac{1}{2}\left\{1 + \cos\left(\frac{r - (r(k) - \delta r)}{2\delta r}\pi\right)\right\} & (r(k) - \delta r \leq r \leq r(k) + \delta r) \\ 0 & (r > r(k) + \delta r) \end{cases} \tag{9}$$

where $\delta r = 0.25$ Å. The number of residues located in the shell between the sphere of radius $r(k-1)$ and that of radius $r(k)$ is denoted by $N_k^{j+4}$, which is calculated as

$$\begin{aligned} N_k^{j+4} &= \sum_{n \neq j+4}^{N^{res}} u_k(r_{j+4,n}) - \sum_{m \neq j+4}^{N^{res}} \Delta_k(r_{j+4,m}) & (2 \leq k \leq 4) \\ N_k^{j+4} &= \sum_{n \neq j+4}^{N^{res}} u_k(r_{j+4,n}) & (k = 1) \end{aligned}, \tag{10}$$

where $\Delta_k(r) = 1$ for $r < r(k-1)$ and $\Delta_k(r) = 0$ for $r \geq r(k-1)$. The constraint to $N_k^{j+4}$ is estimated by sampling $N_k^{5,i}$ which is the number of neighboring residues around the center residue of $F_i(j)$ and is represented in the energy term as

$$V(N_k^{j+4}) = \frac{-1}{\sum_{i=1}^{N^{\text{fragment}}} C(i,j)} \sum_{i=1}^{N^{\text{fragment}}} C(i,j) \times \exp\left[\frac{-\left(N_k^{j+4} - N_k^{5,i}\right)^2}{2C_{\text{nei}}}\right], \quad (11)$$

where $C_{\text{nei}} = 0.5$ and $N^{\text{fragment}} = 40$. $V_{\text{nn}}$ is defined by summing $V(N_k^{j+4})$ for $k$ and for all the 9-residue windows in the target;

$$V_{\text{nn}} = \sum_{j=1}^{N^{\text{res}}-8} \sum_{k=1}^{4} V(N_k^{j+4}). \quad (12)$$

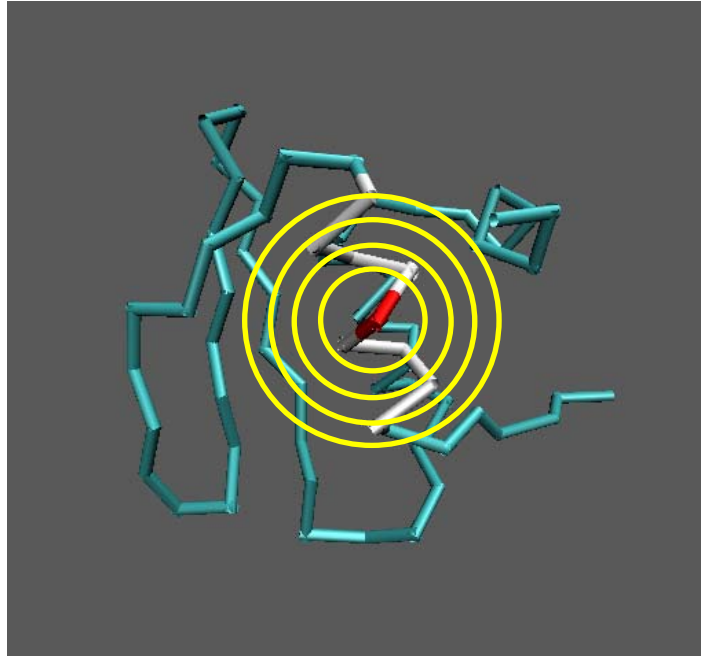

Figure 2
Illustration to explain the construction of $V_{\text{nn}}$. $V_{\text{nn}}$ gives constraints for the number of residues residing in layers between spheres around each residue.

**Beta-sheet potential.** $V_\beta$ represents stabilization of β-sheet through the formation of pseudo-hydrogen bonds among β-strands. First, by defining a vector along the chain, $\mathbf{a}_n = \mathbf{r}_{n+1,n+2}/r_{n+1,n+2}$, and by using $l_0 = \cos(\pi/3)$, $\Delta l = 0.3$, $\theta_0 = (1+10/180)\pi$, and $\Delta c = 0.25$, formation of the β-strand is expressed by a function,

$$V_{\text{strand}}(n) = \exp\left(-\frac{(\cos(\theta_n - \theta_0) - 1)^2}{2\Delta c^2}\right) \times \exp\left(-\frac{1}{2\Delta l^2}\left[(\vec{a}_{n-1}\cdot\vec{a}_n - l_0)^2 + (\vec{a}_n\cdot\vec{a}_{n+1} - l_0)^2\right]\right). \quad (13)$$

The geometrical constraint for the formation of pseudo-hydrogen bonds is expressed by a function,

$$V_{\text{bond}}^{k,l}(n,m) = \exp\left(-\frac{(r_{n+k,m+l} - l_{hb})^2}{2\Delta l_{hb}^2}\right)\exp\left(-\frac{1}{2\Delta s^2}\left[(\mathbf{a}_n\cdot\mathbf{b}_{n+k,m+l})^2 + (\mathbf{a}_m\cdot\mathbf{b}_{n+k,m+l})^2\right]\right), \quad (14)$$

where $\mathbf{b}_{n,m} = \mathbf{r}_{n,m}/r_{n,m}$ is a vector directing parallel to hydrogen bonds, and $\Delta s = 0.3$, $\Delta l_{hb} = 1.0\,\text{Å}$, and $l_{hb} = 5.0\,\text{Å}$. The parallel and anti-parallel association of the β-strand around the *n*th residue and the β-strand around the *m*th residue are expressed by functions,

$$V_{\text{parallel}}(n,m) = w_{n+1,m+1}V_{\text{bond}}^{1,1}(n,m) + w_{n+2,m+2}V_{\text{bond}}^{2,2}(n,m) + w_0 V_{\text{bond}}^{1,1}(n,m)V_{\text{bond}}^{2,2}(n,m), \quad (15)$$

$$V_{\text{anti-parallel}}(n,m) = w_{n+1,m+2}V_{\text{bond}}^{1,2}(n,m) + w_{n+2,m+1}V_{\text{bond}}^{2,1}(n,m) + w_0 V_{\text{bond}}^{1,2}(n,m)V_{\text{bond}}^{2,1}(n,m), \quad (16)$$

where the coefficient $w_0 = 0.5$, and the coefficients $w_{n,m}$ are determined by using BETApro [27], which is the algorithm based on the neural-network estimation of the probability of the β-sheet pairing: $w_{n,m}$ is defined by

$$w_{n,m} = w_{bp}(n,m) + 0.5, \qquad (17)$$

where $w_{bp}(n,m)$ is the pseudo-energy calculated by BETApro, which reflects the probability of the β-sheet pairing between the strand containing the $n$th residue and the strand containing the $m$th residue. If the $n$th or $m$th residue is not included in the strands predicted by BETApro, $w_{bp}(n, m)$ is set to 0. Using Eqs.13-17, $V_\beta$ is defined by

$$V_\beta = -\sum_{n=1}^{N^{\text{res}}-7}\sum_{m=n+4}^{N^{\text{res}}-3} V_{\text{strand}}(n) V_{\text{strand}}(m) \times \left(V_{\text{parallel}}(n,m) + V_{\text{anti-parallel}}(n,m)\right). \qquad (18)$$

**Langevin MD simulation.** Using Eqs.4, 8, 12, and 18, $V_{\text{total}}$ of Eq.2 can be analytically differentiated by $\{\mathbf{r}_i\}$. Weight factors $w_1$, $w_2$, $w_3$, and $w_4$ are chosen to give a large enough energy gap between the target structure and the uncorrelated compact structures. We here use $w_1 = 1.5$, $w_2 = 0.75$, $w_3 = 0.79$, and $w_4 = 1.0$. Starting from a stretched chain configuration, the Langevin MD of Eq.1 is performed to search low energy structures with the annealing schedule explained in *Supplementary Text*.

**RESULTS AND DISCUSSION**
For the benchmarking test, the medium and hard targets were selected from categories of TBM, FM, and TBM/FM (the boundary category between TBM and FM) of CASP7. In order to make this benchmarking test sufficiently stringent, we used SSEARCH program [28] with E-value 10.0 and PSI-BLAST program [26] with E-value 0.01 to exclude the homologous proteins to targets from the protein structure library to construct $V_{\text{fragment}}^{\text{pair}}$, $V_{\text{fragment}}^{\text{angle}}$ and $V_{\text{nn}}$. Results of the test are summarized in Table 1 in terms of the Global Distance Test Total Score (GDT_TS) [29]. The column of "KORO-0" in Table 1 represents the results of our participation in CASP7: The present authors have participated in CASP7 with the team name "KORO" by using the method explained in this paper [30]. In KORO-0, we repeated the Langevin MD calculation $N_{\text{traj}}$ = 100-400 times for each target by using different random number seeds. From $N_{\text{traj}}$ structures obtained at the last step of $N_{\text{traj}}$ trajectories, the lowest and second-lowest energy structures were selected as the 1st and 2nd models. We also performed the cluster analysis of $N_{\text{traj}}$ structures and the center structures of the 1st, 2nd, and 3rd largest clusters were selected as the 3rd, 4th, and 5th models. Results shown in Table 1 are GDT_TS of the 1st model and the best GDT_TS of five models. Also shown in column of KORO-1 in Table 1 are the results re-calculated with the fixed number of $N_{\text{traj}}$ = 400.

As shown in Table 1, the method provides considerably high values of GDT_TS for some targets. T0283 and T0354 are examples of such targets. For those targets, energy of $N_{\text{traj}}$ = 400 structures and their GDT_TS are plotted in Figure 3. Structures of the 1st model and the experimentally observed structures are compared in Figure 4. The absolute value of the correlation coefficient, $|C_{\text{Energy-GDTTS}}|$, is 0.340 (p-value < $10^{-6}$) for T0283, and 0.556 (p-value < $10^{-6}$) for T0354. Fairly large values of $|C_{\text{Energy-GDTTS}}|$ for these targets imply that the energy surfaces for these targets shape funnel-like landscapes. Also shown in Figures 3 and 4 are the plot and structures for T0300. For T0300, $|C_{\text{Energy-GDTTS}}|$ is as low as 0.096 (p-value 0.056). The weak correlation for T0300 implies that the present energy function misses some important features to characterize the energy landscape of this protein. GDT_TS of the lowest energy structure for T0300 is relatively low as shown in Table 1. Among $N_{\text{traj}}$ structures, however, we can find ones with fairly large GDT_TS, so that the one way to rescue those good structures is to use a different score function which can discriminate the candidates from other structures generated by Langevin MD. For this purpose, we use an empirical score-function whose derivation is explained in *Supplementary Text*.

First, the score-function is applied to $N_{\text{traj}}$ = 400 structures and $N_{\text{score}}$ structures which have the highest score are selected from $N_{\text{traj}}$ structures. For 14 among 18 targets, more refined structures are obtained with the evident increase in $|C_{\text{Energy-GDTTS}}|$ by limiting candidates from 400 to $N_{\text{score}}$ = 50 structures (see *Supplementary Figure*). Hence, the better results are expected by this cross-checking to use energy and score at the same time. In Figure 3, we show the plots for $N_{\text{score}}$ = 50 structures. By choosing $N_{\text{score}}$ = 50 structures, $|C_{\text{Energy-GDTTS}}|$ for T0283, T0354, and T0300 is 0.434 (p-value 0.0016), 0.591 (p-value 6.3X$10^{-6}$), and 0.384 (p-value 0.006). In Figure 4, we can find a substantial improvement in the predicted structure of T0300. Thus, the problem of small $|C_{\text{Energy-GDTTS}}|$ for T0300 is resolved by introducing this score function.

In KORO-2, we use $N_{score} = 50$ and select the lowest energy structure of $N_{score}$ structures as the 1st model, the lowest energy structure of $N_{traj}$ structures as the 2nd model and center structures of the 1st, 2nd, and 3rd largest clusters of $N_{traj}$ structures as 3rd, 4th and 5th models. Results are summarized in the column of KORO-2 of Table 1. In KORO-2, the 1st models are improved from KORO-1 for 10 targets, show no change for 3 targets, and become to have the smaller GDT_TS for 5 targets. The best of five models are improved for 4 targets, show no change for 11 targets, and become to have the smaller GDT_TS for 3 targets. Thus, we can find that the results are overall improved from KORO-1.

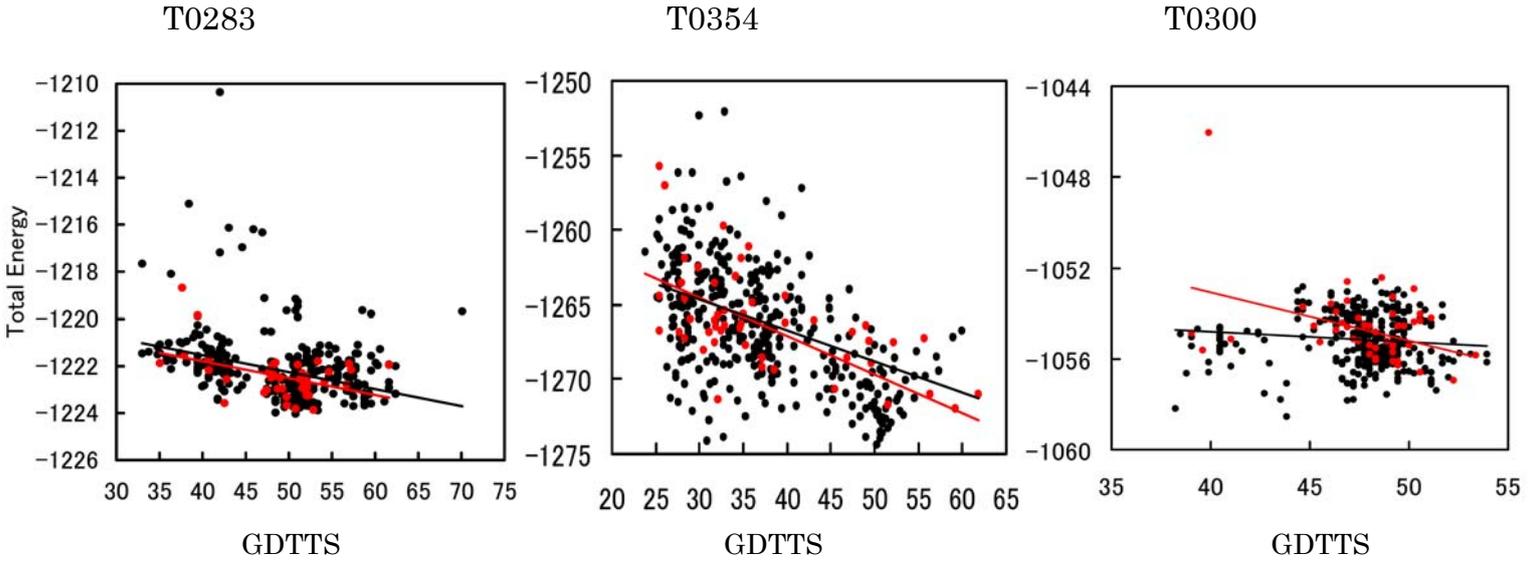

Figure 3
Plot of GDT_TS vs. total energy. Red dots are $N_{score} = 50$ structures of highest score and black dots are other $N_{traj} - N_{score} = 350$ structures. Gradient of red and black lines represents the correlation coefficient of red $N_{score}$ dots and all of $N_{traj}$ dots, respectively. Left: T0283, Middle: T0354, and Right: T0300.

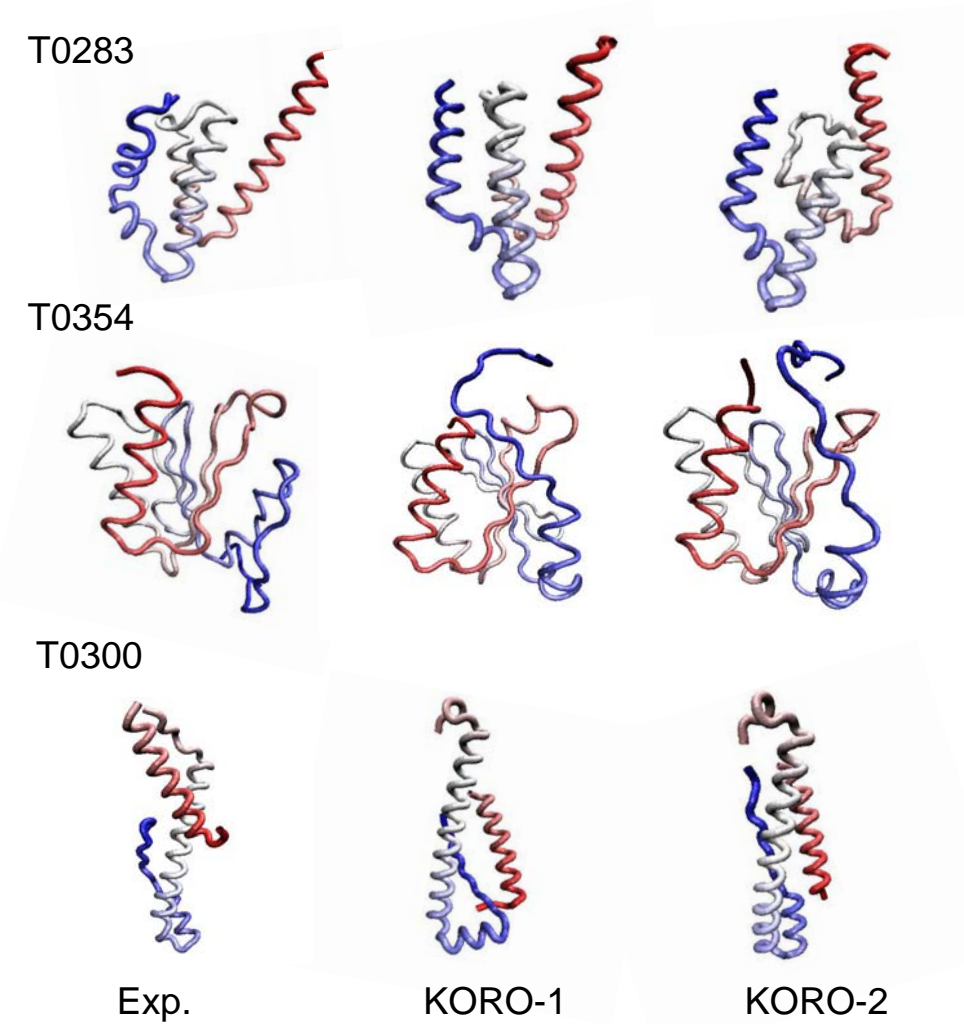

Figure 4

Comparison of structures. Chains are colored from red (N terminus) to blue (C terminus). **Top**: Structures for T0283. (left) The experimentally observed structure of PDB code 2HH6, where the residues 1-97 used for the evaluation in CASP7 are shown, (middle) the 1st model of KORO-1 with GDT_TS = 50.77 and RMSD (root mean square deviation) = 5.19Å from 2HH6 for residues 1-97, and (right) the 1st model of KORO-2 with GDT_TS = 52.84 and RMSD = 5.71Å. **Middle**: Structures for T0354. (left) The experimentally observed structure of PDB code 2ID1, (middle) the 1st model of KORO-1 with GDT_TS = 50.21 and RMSD = 3.23Å and (right) the 1st model of KORO-2 with GDT_TS = 59.17 and RMSD = 4.30Å. Here, RMSD is calculated by using residues 1-100 (by disregarding residues 101-120). **Bottom**: Structures for T0300. (left) The experimentally observed structure of PDB code 2H3R, where structure from 32nd to 38th residue has not been determined experimentally, (middle) the 1st model of KORO-1 with GDT_TS = 43.82 and RMSD = 12.28Å, and (right) the 1st model of KORO-2 with GDT_TS = 52.25 and RMSD = 10.77Å.

In Tables 2 and 3, GDT_TS of KORO-2 is compared with the results of other groups participated in CASP7: Compared are ROKKO, TASSER, Baker, and Zhang [30]. Both GDT_TS of the 1st model (Table 2) and GDT_TS of the best of five models (Table 3) show that KORO-2 achieved results comparable with other approaches. The combined use of energy and score improved the results as the number of targets showing the highest GDT_TS of five different approaches for 18 targets were 2 for the 1st model and 3 for the best of five models when KORO-1 is compared with other four approaches and 6 for the 1st model and 4 for the best of five models when KORO-2 is used as in Tables 2 and 3.

As shown in Tables 2 and 3, all approaches hitherto developed are still *not* satisfactory for providing high enough GDT_TS results for many FM targets consistently. It is, therefore, strongly desired to test new ideas aiming for a more consistent prediction. In this paper, we proposed a new method of *de novo* structure prediction by simulating the folding process with the Langevin MD calculation. The benchmarking test showed that the results are further improved by cross-checking structures with two criteria of energy and score.

It is evident that there is a large room for improvement in the present model. For example, representation of the chain conformation should be refined with the more detailed degrees of freedom, and use of the variable length local structures instead of the fixed 9-residue fragments may help to search the consistent structures more efficiently. The method developed in this paper showed that the dynamical searching of structures satisfying the local and global multi-residue constraints defined through the sequence profile analyses should be a way to proceed toward a more consistent method of *de novo* prediction.

**Acknowledgement**

This work was supported by grants from the Ministry of Education, Culture, Sports, Science, and Technology, Japan, and by grants for the 21st century COE program for Frontiers of Computational Science.

**Table 1.** GDT_TS of the first model structure and GDT_TS of the best of five model structures obtained by the CASP7 participation (KORO-0), re-calculation after CASP7 (KORO-1), and the combined use of energy and score (KORO-2).

| Target | Category | Type | NT$^\dagger$ | KORO-0* | KORO-1* | KORO-2* |
|---|---|---|---|---|---|---|
| T0283 | TBM | α | 97 | 50.00(62.88) | 50.77(62.89) | 52.84(62.89) |
| T0287 | FM | α | 161 | 28.11(28.11) | 28.11(28.11) | 24.53(28.11) |
| T0300 | FM | α | 89 | 40.17(52.25) | 43.82(49.44) | 52.25(52.25) |
| T0304 | TBM/FM | α+β | 101 | 30.69(34.16) | 30.45(35.40) | 33.66(35.40) |
| T0307 | FM | α | 123 | 21.55(21.55) | 19.31(25.81) | 22.15(25.81) |
| T0309 | FM | β | 62 | 30.65(31.85) | 31.45(33.47) | 28.23(33.47) |
| T0314 | FM | α+β | 103 | 19.66(26.21) | 22.57(23.06) | 22.57(22.57) |
| T0316_2 | FM | β | 60 | *no data* | 42.50(42.50) | 34.17(42.50) |
| T0319 | FM | α+β | 135 | 27.59(27.59) | 20.00(22.22) | 22.41(22.41) |
| T0321_2 | TBM/FM | α+β | 148 | 29.39(29.39) | 41.05(41.05) | 32.43(41.05) |
| T0347_2 | FM | α | 71 | 33.80(38.38) | 35.56(38.03) | 35.36(38.03) |
| T0348 | TBM/FM | α+β | 61 | 49.59(52.87) | 45.90(50.82) | 44.67(50.82) |
| T0350 | FM | α+β | 91 | 30.49(31.32) | 29.40(31.04) | 30.50(31.04) |
| T0353 | FM | α+β | 83 | 28.92(29.22) | 31.63(42.77) | 31.63(31.63) |
| T0354 | TBM | α+β | 120 | 51.23(51.23) | 50.21(50.21) | 59.17(59.17) |
| T0356_3 | FM | α+β | 120 | 18.12(22.92) | 26.88(29.58) | 31.04(31.04) |
| T0361 | FM | α | 158 | 32.12(32.12) | 27.85(32.60) | 32.12(32.12) |
| T0382 | TBM/FM | α | 119 | 30.46(39.08) | 30.25(33.82) | 33.82(33.82) |

* In each column, results of the 1st model are shown first and those of the best of five models are shown in parentheses.

$^\dagger$ Number of residues possible to evaluate.

**Table 2.** Comparison of GDT_TS of the 1st models obtained by 5 different approaches.

| Target | KORO-2 | ROKKO | TASSER | Baker | Zhang |
|---|---|---|---|---|---|
| T0283 | **62.89 | 45.36 | 60.05 | *75.00 | 40.47 |
| T0287 | *28.11 | 16.30 | **22.36 | 19.72 | 22.05 |
| T0300 | *52.25 | *no data* | 38.20 | **44.10 | 40.45 |
| T0304 | 35.40 | 22.52 | 28.96 | *45.05 | **44.55 |
| T0307 | 25.81 | **28.05 | 24.59 | *30.08 | 27.24 |
| T0309 | 33.47 | *36.69 | 31.05 | **34.27 | 30.24 |
| T0314 | 22.57 | 21.12 | **23.79 | 23.06 | *24.52 |
| T0316_2 | *42.50 | 20.00 | 22.91 | **31.67 | 31.25 |
| T0319 | **22.41 | 17.78 | 18.33 | *25.74 | 19.81 |
| T0321_2 | 41.05 | 35.98 | **45.77 | 42.91 | *45.95 |
| T0347_2 | *38.03 | 36.27 | 33.45 | 32.40 | **35.56 |
| T0348 | 45.08 | 30.33 | **47.54 | 45.90 | *50.00 |
| T0350 | 31.04 | 32.42 | 35.44 | **57.42 | *57.97 |
| T0353 | 31.63 | 40.36 | 37.05 | **40.97 | *51.51 |
| T0354 | *59.17 | 42.21 | 44.67 | **56.76 | 43.03 |
| T0356_3 | **31.67 | 18.75 | 31.04 | 22.09 | *33.12 |
| T0361 | *32.12 | 18.04 | 27.85 | **29.43 | 19.62 |
| T0382 | 33.82 | **47.69 | 46.22 | 38.23 | *50.21 |

\* The highest GDT_TS of five approaches

\*\* The 2nd highest GDT_TS of five approaches

**Table 3.** Comparison of GDT_TS of the best of five models obtained by 5 different approaches.

| Target | KORO-2 | ROKKO | TASSER | Baker | Zhang |
|--------|--------|-------|--------|-------|-------|
| T0283 | **62.89 | 45.36 | 60.05 | *82.47 | 58.76 |
| T0287 | *28.11 | 21.59 | 24.53 | 23.75 | *28.11 |
| T0300 | **52.25 | *no data* | 38.20 | *53.37 | 46.07 |
| T0304 | 35.40 | 42.08 | 31.19 | *45.05 | **44.55 |
| T0307 | 25.81 | **34.96 | 24.59 | *39.02 | 30.69 |
| T0309 | 33.47 | *36.69 | 33.06 | **34.27 | 32.26 |
| T0314 | 22.57 | **25.00 | 24.27 | 23.06 | *29.13 |
| T0316_2 | *42.50 | 20.84 | 25.00 | 34.58 | **34.59 |
| T0319 | 22.41 | 20.00 | 22.59 | *30.92 | **25.37 |
| T0321_2 | 41.05 | 37.67 | **45.77 | 45.44 | *45.95 |
| T0347_2 | 38.03 | **41.90 | 37.67 | *52.82 | 39.08 |
| T0348 | 50.82 | 48.77 | 49.18 | **53.28 | *54.51 |
| T0350 | 31.04 | 43.41 | 35.44 | *60.16 | **57.97 |
| T0353 | 31.63 | **46.99 | 44.28 | 44.58 | *51.51 |
| T0354 | *59.17 | 42.21 | 47.13 | **57.79 | 52.46 |
| T0356_3 | 31.04 | 30.00 | *33.96 | 27.71 | **33.12 |
| T0361 | *32.12 | 18.67 | 27.85 | **31.01 | 28.64 |
| T0382 | 33.82 | 47.69 | **53.36 | 51.05 | *59.66 |

\* The highest GDT_TS of five approaches
\*\* The 2nd highest GDT_TS of five approache

**References**

[1] K. Joo, J. Lee, S. Lee, J.-H. Seo, S. J. Lee, and J. Lee, High accuracy template based modeling by global optimization, Proteins 69 Suppl.8 (2007) 83-89.

[2] R. Jauch, H. C. Yeo, P. R. Kolatkar, and N. D. Clarke, Assessment of CASP7 structure predictions for template free targets, Proteins 69 Suppl.8 (2007) 57-67.

[3] A. Tramontano, Worth the effort: an account of the Seventh meeting of the worldwide critical assessment of Techniques for Protein Structure Prediction, FEBS J. 274 (2007) 1651-1654.

[4] C. A. Rohl, C. E. M. Strauss, K. M. S. Misura, and D. Baker, Protein structure prediction using Rosetta, Methods Enzymol. 383 (2004) 66-93.

[5] P. Bradley, K. M. S. Misura, and D. Baker, Toward high-resolution de novo structure prediction for small proteins, Science 309 (2005) 1868-1871.

[6] J. Lee, S.-Y. Kim, and J. Lee, Protein structure prediction based on fragment assembly and parameter optimization, Biophys. Chem. 115 (2005) 209-214.

[7] Y. Fujitsuka, G. Chikenji, and S. Takada, SimFold energy function for de novo protein structure prediction: consensus with Rosetta, Proteins 62 (2006) 381-398.

[8] T. Ishida, T. Nishimura, M. Nozaki, T. Terada, S. Nakamura, and K. Shimizu, Development of an ab initio protein structure prediction system ABLE, Genome Inform. 14 (2003) 228-237.

[9] Y. Zhang, A. K. Arakaki, and J. Skolnick, TASSER: an automated method for the prediction of protein tertiary structures in CASP6, Proteins. 61 Suppl.7 (2005) 91-98.

[10] H. Zhou and J. Skolnick, Ab initio protein structure prediction using Chunk-TASSER, Biophys. J. 93 (2007) 1510-1518.

[11] S. Wu, J. Skolnick, and Y. Zhang, Ab initio modeling of small proteins by iterative TASSER simulations, BMC Biol. 5:17 (2007).

[12] H. Zhou, S. B. Pandit, S. Y. Lee, J. Borreguero, H. Chen, L. Wroblewska, and J. Skolnick, Analysis of TASSER-based CASP7 protein structure prediction results, Proteins. 69 Suppl.8 (2007) 90-97.

[13] C.B. Anfinsen, Principles that govern the folding of protein chains, Science 181 (1973) 223-230

[14] N. Go, Theoretical studies of protein folding, Ann. Rev. Biophys. Bioeng. 12 (1983) 183-210.

[15] J.N. Onuchic, Z. Luthey-Schulten, and P.G. Wolynes, Theory of protein folding: the energy landscape perspective, Ann. Rev. Phys. Chem. 48 (1997) 545-600.

[16] G. Chikenji, Y. Fujitsuka, and S. Takada, Shaping up the protein folding funnel by local interaction: lesson from a structure prediction study, Proc. Natl. Acad. Sci. USA

103 (2006) 3141-3146.

[17] R. Kazmierkiewicz, A. Liwo, and H. A. Scheraga, Energy-based reconstruction of a protein backbone from its alpha-carbon trace by a Monte-Carlo method, J. Comput. Chem. 23 (2002) 715-723.

[18] M. Nanias, M Chinchio, S Oldziej, C. Czaplewski, and H. A. Sheraga, Protein structure prediction with the UNRES force-field using replica-exchange monte carlo-with-minimization; comparison with MCM, CSA and CFMC, J. Comput. Chem. 26 (2005) 1472-1486.

[19] A. Liwo, M. Khalili, and H. A. Scheraga, Ab initio simulations of protein-folding pathways by molecular dynamics with the united-residue model of polypeptide chains, Proc. Natl. Acad. Sci. USA 102 (2005) 2362-2367.

[20] G. A. Papoian, J. Ulander, M. P. Eastwood, Z. Luthey-Schulten, P. G. Wolynes, Water in protein structure prediction, Proc. Natl. Acad. Sci. USA 101 (2004) 3352-3357.

[21] C. Hardin, M. P. Eastwood, Z. Luthey-Schulten, and P. G. Wolynes, Associative memory hamiltonians for structure prediction without homology: alpha-helical proteins, Proc. Natl. Acad. Sci. USA 97 (2000) 14235-14240.

[22] T. N. Sasaki, and M. Sasai, A coarse-grained Langevin molecular dynamics approach to protein structure reproduction, Chem. Phys. Lett. 402 (2005) 102-106.

[23] G. Wang, and R. L. Dunbrack, Jr, PISCES: a protein sequence culling server, Bioinformatics 19 (2003) 1589-1591.

[24] http://dunbrack.fccc.edu/PISCES.php

[25] http://www.ncbi.nlm.nih.gov/

[26] S. F. Altschul, T. L. Madden, A. A. Schaffer, J. Zhang, Z. Zhang, W. Miller, and D. J. Lipman, Gapped BLAST and PSI-BLAST: a new generation of protein database search programs, Nucleic. Acids. Res. 25 (1997) 3389-3402.

[27] J. Cheng, and P. Baldi, Three-stage prediction of protein beta-sheets by neural networks, alignments and graph algorithms, Bioinformatics 21, Suppl.1 (2005) i75-i84.

[28] W. R. Pearson, Searching protein sequence libraries: Comparison of the sensitivity and selectivity of the Smith-Waterman and FASTA algorithms, Genomics 11 (1991) 635-650.

[29] A. Zemla, LGA: a method for finding 3D similarities in protein structures, Nucleic Acids Res. 31 (2003) 3370-3374.

[30] http://www.predictioncenter.org/casp/casp7/public/cgi-bin/results.cgi

# Supplementary Text

## A coarse-grained Langevin molecular dynamics approach to *de novo* protein structure prediction

Takeshi N. Sasaki, Hikmet Cetin and Masaki Sasai

**The annealing schedule.** Langevin molecular dynamics (MD) was performed with $T(i) = T_0(1-\text{Int}(i/N_{\text{ann}})/(N_{\text{step}}/N_{\text{ann}}))$, where $T(i)$ is $T$ at the $i$th step of MD, $\text{Int}(x)$ is the largest integer smaller than $x$, and $T_0=0.3$. $N_{\text{step}} = N^{\text{res}} \times 10^6$ is the number of total steps of MD, and $N_{\text{ann}}$ is set to 10.

**Evaluation of score.** Structures are characterized by three indices. (1) Secondary structure, S; The dihedral angle $\eta_i$ between the plane spanned by $\mathbf{r}_{i-1}$, $\mathbf{r}_i$, and $\mathbf{r}_{i+1}$, and the plane spanned by $\mathbf{r}_i$, $\mathbf{r}_{i+1}$, and $\mathbf{r}_{i+2}$ are defined at the $i$th residue of the structure. The local structure around the $i$th site is classified into S = α when $0.35 < \cos\eta_i$, S = β when $\cos\eta_i < -0.8$, and S = C when $-0.8 \leq \cos\eta_i \leq 0.35$. (2) Local density, $N_{10}$; The number of α-carbons in the sphere of radius 10Å around the position $\mathbf{r}_i$. (3) The local contact order, LCO; Average of the distance along the sequence, $|i - j|$, between the centered $i$th residue and the other $j$th residue located in the sphere of radius 10Å around the position $\mathbf{r}_i$. In this way, the local structure around each residue is represented by (S, $N_{10}$, LCO).

Using thus defined set of three indices, the local structure around each residue is classified into 24 classes, class A to class X, as in Table 1. Then, the protein structure is represented by a sequence of classes $\{x_i\}$ as $x_1 = B$, $x_2 = B$, $x_3 = X$, ..., $x_N = A$, for example. Each of $N_{\text{traj}}$ structures generated by Langevin MD is represented by such a sequence of classes.

The score function is constructed for each target protein. Using the same library as was used in Text for the fragment selection, the 9-residue fragments which have the correlation coefficient larger than 0.6 to the 9-residue window in the target are selected. We define $N_i^{0.6}$ as the number of fragments thus selected for the 9-residue window around the $i$th residue in the target. $N_i^{0.6}$ fragments are sorted in order of the correlation coefficient, and the weight of the $k$th fragment, $C_i(k)$, is defined by $C_i(k) = (N_i^{0.6} - O_i(k) + 1)/N_i^{0.6}$, where $O_i(k)$ is the order of the $k$th fragment in the result of sorting. We also introduce an index $p_k(y)$: $p_k(y) = 1$ when the center residue of the $k$th fragments has a class $y$ and $p_k(y) = 0$, otherwise. Then, the score function for the $i$th residue to take class $x_i$, $Score_i(x_i)$, is defined by $Score_i(x_i) = \sum_k C_i(k) p_k(x_i)$, where the summation over $k$ is a sum over $N_i^{0.6}$ fragments. The score of the structure generated by Langevin MD is evaluated as $Score = \sum_i Score_i(x_i)$.

**Table 1.** Definition of classes from A to X.

|  | $N_{10} \leq 6$ | $7 \leq N_{10} \leq 12$ | $13 \leq N_{10} \leq 18$ | $19 \leq N_{10}$ |  |
|---|---|---|---|---|---|
| $S = \alpha$ | A | B | C | D | LCO < 27.79 |
| $S = \beta$ | E | F | G | H |  |
| $S = C$ | I | J | K | L |  |
| $S = \alpha$ | M | N | O | P | LCO $\geq$ 27.79 |
| $S = \beta$ | Q | R | S | T |  |
| $S = C$ | U | V | W | X |  |

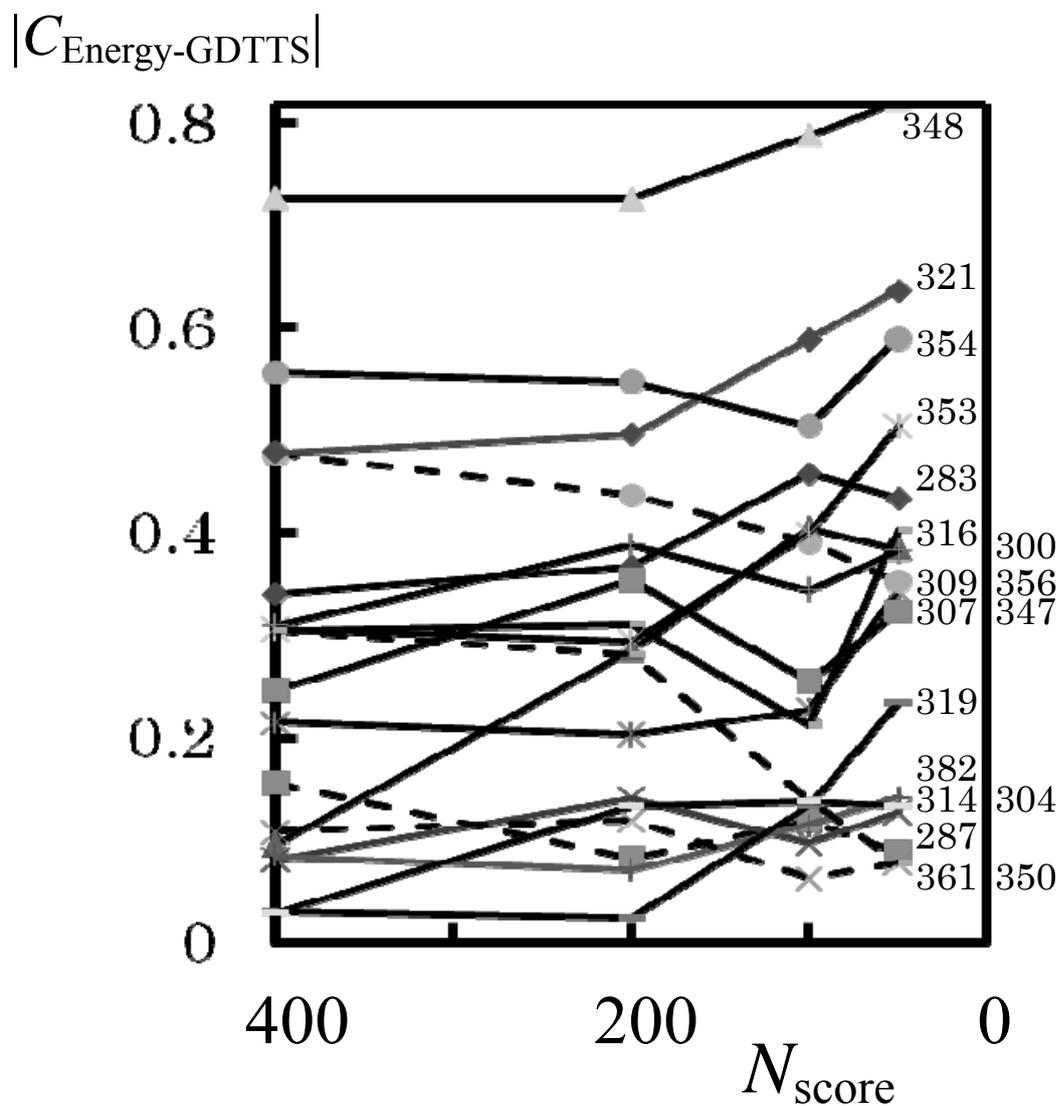

**Supplementary Figure 1**

Dependence of the absolute value of the correlation coefficient between GDT_TS and total energy, $|C_{\text{Energy-GDTTS}}|$, on the number of selected structures by using the score function, $N_{\text{score}}$. For 14 target proteins, $|C_{\text{Energy-GDTTS}}|$ for $N_{\text{score}} = 50$ is larger than that for $N_{\text{score}} = 400$ (real lines) but for 4 target proteins $|C_{\text{Energy-GDTTS}}|$ decreases as $N_{\text{score}}$ decreases (dashed lines).